# New Tools for Low Energy Dynamical Supersymmetry Breaking

Michael Dine[a], Ann E. Nelson[b], Yosef Nir[c] and Yuri Shirman[a]

[a]*Santa Cruz Institute for Particle Physics, University of California, Santa Cruz, CA 95064*
[b]*Department of Physics, Box 351560, University of Washington, Seattle, WA 98195-1560*
[c]*Department of Particle Physics, Weizmann Institute of Science, Rehovot 76100, ISRAEL*

## Abstract

We report the construction of large new classes of models which break supersymmetry dynamically. We then turn to model building. Two of the principal obstacles to constructing simple models of dynamical supersymmetry breaking are the appearance of Fayet-Iliopoulos D terms and difficulties in generating a $\mu$ term for the Higgs fields. Among the new models are examples in which symmetries prevent the appearance of Fayet-Iliopoulos terms. A gauge singlet field, that may play a role in explaining the hierarchy in quark and lepton parameters, can generate a suitable $\mu$ term. The result is a comparatively simple model, with a low energy structure similar to that of the MSSM, but with far fewer arbitrary parameters. We begin the study of the phenomenology of these models.

# 1. Introduction: Survey of Schemes for Dynamical Supersymmetry Breaking

If supersymmetry plays a role in low energy physics, it is presumably dynamically broken. In this paper we greatly extend the list of gauge theories which are known to dynamically break supersymmetry; readers who are only interested in these new examples should just read §2 and the Appendix. In §3 we show how to build reasonably simple realistic models, where supersymmetry is dynamically broken at low energies. These have the light particle content of the usual minimal supersymmetric standard model but are much more predictive, with fewer necessary assumptions. In §4 we discuss the phenomenology of an example where all the masses of undiscovered particles lighter than a TeV may be predicted in terms of just two new parameters. In §5 we remind the reader why the problem of Higgs doublet-triplet splitting in supersymmetric grand unified theories is more easily solved with low energy supersymmetry breaking. We review some cosmological issues in §6. In the remainder of this introduction we discuss some problems of existing theories in which supersymmetry is dynamically broken, and how several new tools help us construct better theories.

There are various ways dynamical supersymmetry breaking might arise. In theories like string theory, there are classically many flat directions in the potential. Such flat directions are often lifted by non-perturbative effects [1]. Typically the potentials which are generated in these flat directions fall to zero for large values of the fields. The most familiar and notorious example of this kind is the dilaton of string theory, whose potential always tends to zero at weak coupling [2]. Such potentials might be stabilized by multiple condensates [3], or perhaps more plausibly by large corrections to the Kahler potential in the strong coupling region [4]. It is fair to say that no very compelling model of the first type exists. The second scheme is basically a hope; it is unlikely that any explicit computation will verify such a picture soon.

Even if such schemes are successful, there are many issues which such models have to face. Among these is the question of flavor changing neutral currents. Solving this problem in the framework of supersymmetry requires either a high degree of squark degeneracy or alignment of quark and squark mass matrices [5]. Some suggestions for achieving squark degeneracy in the string context exist [6-7], but they require that string theory be truly weakly coupled, in the sense that perturbation theory should be good for the Kahler potential. It is hard to see how this can be consistent with the expected behavior of the dilaton potential. Alternatively, non-Abelian flavor symmetries may play some role [8]. Other issues include a variety of cosmological problems, perhaps the most severe being the moduli problem [9]. One solution to the latter problem is that the dilaton and moduli are stabilized by nonperturbative physics at high energies and play no role in the breaking of supersymmetry [9]; another possible solution is weak scale inflation [10,11].

Alternatively, models are known in which supersymmetry is broken without flat directions [1]. In such cases, one does not require the intervention of complicated stabilization mechanisms. As in the case of flat directions described above, one can imagine breaking supersymmetry at a scale intermediate between $M_W$ and $M_p$. This idea, however, turns out to be fraught with difficulties, particularly with obtaining appreciable gluino masses [1,9,12]. Alternatively, one can imagine breaking supersymmetry at comparatively low energies, of order 10's–1000's of TeV. In this case, gauge interactions can serve as the "messengers" of supersymmetry breaking. Apart from the fact that the physics of supersymmetry breaking is potentially accessible, such a scheme has an immediate bonus: there is automatically sufficient squark and slepton degeneracy to understand the absence of flavor changing neutral currents.

Early efforts to build models along these lines suffered from a number of difficulties. The most severe of these were that $SU(3)_C$ typically became strong a few decades above the scale of supersymmetry breaking, and that the known models all possessed (astrophysically) dangerous light axions from a spontaneously broken R symmetry. Nelson and Seiberg noted that dimension five operators expected from Planck scale physics could explicitly break the R symmetry and give the axion a sufficiently large mass so that it would not be produced in stars [13] without restoring supersymmetry. Bagger, Poppitz and Randall pointed out that when R symmetry and supersymmetry break at the same scale, cancellation of the cosmological constant within the framework of supergravity by adding a constant to the superpotential [14] necessarily requires such explicit R symmetry breaking but also does not restore supersymmetry. Solutions to the first problem were provided in refs. [15,16]. Here it was suggested that a new gauge symmetry, referred to as the "messenger" gauge group, could play a crucial role. These models, while potentially realistic, were fairly complicated. In ref. [16] the messenger group was simply a $U(1)_m$ known as messenger hypercharge. The appearance of Fayet-Iliopoulos D terms for $U(1)_m$ caused a number of problems, forcing several couplings to be extremely small. Also, simple arguments suggested that there could be no $\mu$ term, and extra singlets appeared in the low energy theory, with carefully adjusted couplings, in order to obtain suitable breaking of $SU(2) \times U(1)$.

In the present note, we report substantial progress on these issues. We present new models of dynamical supersymmetry breaking (without flat directions). These significantly extend the known list of such theories, which previously contained just 5 examples [1,17-19]. All of our examples are "calculable" [17], in the sense that by reducing a parameter in the superpotential the supersymmetry breaking scale may be tuned to be small compared with the scale of gauge dynamics and so the ground state may be systematically studied. Using these models, we construct theories without the appearance of messenger group Fayet-Iliopoulos terms and their associated problems. Then, building on an idea of Leurer, Nir and Seiberg [20], we explain how a $\mu$ term of the correct order of magnitude can arise naturally. We finally put these ideas together to construct a model of dynamical supersymmetry breaking which, at low energies, is a version of the MSSM where, once the $Z$ boson mass is fixed, there are only two undetermined parameters. This is in contrast to the usual treatment where, without *ad hoc* assumptions, there are of order $10^2$ unknown parameters. We begin the exploration of the parameter space of this theory, and find that there is a significant region which is presently consistent with all experiments.

## 2. New Models Which Exhibit Dynamical Supersymmetry Breaking

There is a simple criterion for models which exhibit dynamical supersymmetry breaking [1]. If a theory has no flat directions, and it has a global symmetry which is spontaneously broken, then supersymmetry is spontaneously broken. In this section, we describe two new sets of models which satisfy this criterion. One set involves renormalizable interactions only. A second involves non-renormalizable interactions as well.

## 2.1. A renormalizable class of models

In ref. [1], an infinite set of models which break supersymmetry was described. These were models with gauge group $SU(N+4)$, where $N$ was odd, and with $N$ chiral fields, $\bar{F}^a$, in the antifundamental representation and one, $A$, in the antisymmetric tensor representation. Adding the most general superpotential,

$$W = \lambda_{ab} A \bar{F}^a \bar{F}^b, \qquad (2.1)$$

led to a model without flat directions and with a non-anomalous R symmetry. One strategy for constructing generalizations of these models is to take a particular one, and simply discard some of the gauge multiplets while keeping the chiral multiplets. One might then add the most general superpotential allowed in the reduced theory. This procedure is guaranteed to yield chiral models which are free of anomalies. As we will see, the resulting theories often possess non-anomalous R symmetries, and also have no flat directions.

The simplest such model is given by the case $N = 1$, i.e. an $SU(5)$ theory with a $\bar{5}$ and 10. In this case, the superpotential vanishes. One can now modify this theory by taking the gauge group to be the $SU(3) \times SU(2)$ subgroup. Under this group, the $\bar{5}$ and 10 decompose as a $(3, 2)$, two $(\bar{3}, 1)$'s, and a $(1, 2)$. If we add the most general superpotential, we obtain the well-studied $3 - 2$ model of dynamical supersymmetry breaking. We obtain something new if we retain an $SU(4) \times U(1)$ subgroup, where the $U(1)$ generator is

$$Y = \mathrm{diag}(1, 1, 1, 1, -4). \qquad (2.2)$$

The 10 and $\bar{5}$ decompose as an antisymmetric tensor, $A_2$ (the subscript indicates the $U(1)$ charge), a fundamental, $F_{-3}$, an antifundamental $\bar{F}_{-1}$ and a singlet, $S_4$. The most general allowed renormalizable superpotential is

$$W = \lambda S_4 \bar{F}_{-1} F_{-3}. \qquad (2.3)$$

With this superpotential, it is easy to show that there is no flat direction. First note that the most general flat direction of the $SU(4)$ D term has the form

$$A = \begin{pmatrix} a\sigma_2 & 0 \\ 0 & a\sigma_2 \end{pmatrix} \qquad F = \bar{F} = \begin{pmatrix} b \\ 0 \\ 0 \\ 0 \end{pmatrix} \qquad S = c. \qquad (2.4)$$

The $U(1)$ D term requires
$$2|a^2| + 4|c^2| - 4|b|^2 = 0. \tag{2.5}$$
But combined with the vanishing of the F terms, one finds $a = b = c = 0$. In addition to the absence of flat directions, this model also possesses a non-anomalous R symmetry. So one expects that supersymmetry is broken.

To see this in detail, we can ask about the form of the non-perturbative superpotential in the limit that the classical superpotential vanishes. There is, in fact, a unique superpotential consistent with the symmetries:
$$W_{np} = \frac{\Lambda_4^5}{\sqrt{\mathcal{O}}},$$
$$\mathcal{O} = \bar{F}_i F^j A^{ik} A^{lm} \epsilon_{jklm}. \tag{2.6}$$

Even in the presence of the classical superpotential (2.3), symmetry considerations, the known limits $W(\lambda \to 0)$ and $W(\Lambda \to 0)$, and analyticity in $\lambda$ [21,22] still constrain the dynamically generated superpotential to be of the form (2.6).

To see how the term (2.6) is generated, consider first the region of the classical moduli space where $b = c \gg a$. In this direction, $SU(4) \times U(1)$ is broken to $SU(3)$. In the low energy theory, apart from the single light modulus, there is one light 3 and one $\bar{3}$, i.e. one has supersymmetric QCD with one flavor. In this theory, a superpotential is generated non-perturbatively,
$$W_{np} = \frac{\Lambda_3^4}{\sqrt{q\bar{q}}}. \tag{2.7}$$
It is not hard to see that this corresponds precisely to the superpotential above. For example, $\Lambda_3^4 = \Lambda_4^5/b$, so that numerically the superpotentials coincide. In addition, if the $U(1)$ coupling is small, $\lambda \ll g_1 \ll 1$, the low energy theory has approximate flat directions in which $SU(3)$ is broken to $SU(2)$; gluino condensation then generates the required superpotential. Alternatively, one can consider the hierarchy $g_a \ll \lambda \ll 1$. In this case, one expects $a \gg b$. Then at the first stage, the gauge symmetry is broken to $Sp(4) \approx SO(5)$, with two 4's. Again, the appropriate superpotential is generated via gaugino condensation.

To determine the nature of supersymmetry breaking we can minimize the potential in various limits. The simplest case is $\lambda \ll g_a$. Then one expects the minimum to lie in the flat directions of the $SU(4) \times U(1)$ D terms. Rescaling $(a, b, c) = \frac{\Lambda}{\lambda^{1/5}}(a', b', c')$, the scalar potential looks like:

$$V = \lambda^{6/5} \Lambda^4 \left( \left| 2b'c' - \frac{1}{a'b'^2} \right|^2 + |b'|^4 + \left| \frac{1}{a'^4 b'^2} \right| \right). \tag{2.8}$$

The minimum is found at

$$(a, b, c) = \frac{\Lambda}{\lambda^{1/5}}(1.27, 0.97, 0.33), \quad V = 3.3 \times 10^{-4} \lambda^{6/5} \Lambda^4. \tag{2.9}$$

We have also considered the case of small $g_1$. The result above holds reasonably well up to $\lambda \simeq g_1$. For larger values of $\lambda$, there is no simple scaling describing the behaviour of the minimum as a function of $g_1$. Numerically, one finds $a \sim c \gg b$.

*2.2. Generalizations*

There are a vast array of models one can construct in this way. For example, there are a set of models with gauge group $SU(n) \times U(1)$ ($n$ even). Start with the theory with gauge group $SU(n+1)$, an antisymmetric tensor and $n-3$ antifundamentals. Throw out those generators of $SU(n+1)$ which do not lie in an $SU(n) \times U(1)$ subgroup, where the $U(1)$ generator is

$$\tilde{T} = \text{diag}(1, 1, \ldots, 1, -n). \tag{2.10}$$

The original chiral fields decompose as

$$A_2 \ + \ F_{1-n} \ + \ (n-3) \times \bar{F}_{-1} \ + \ (n-3) \times S_n. \tag{2.11}$$

Here $A$ is an antisymmetric tensor, $F(\bar{F})$ is the (anti)fundamental and $S$ a singlet of the $SU(n)$. At the classical level, one can add to this model a superpotential,

$$W = \gamma_{ab} A \bar{F}^a \bar{F}^b + \lambda_{ab} F \bar{F}^a S^b. \tag{2.12}$$

It is not hard to check that for general matrices $\gamma$ and $\lambda$, there are no flat directions; on the other hand, there is a non-anomalous R symmetry, and supersymmetry is broken. To see that there are no flat directions, let us simplify things a bit by taking $\lambda_{ab}$ diagonal.

(i) Suppose first that $F \neq 0$. Then the $\frac{\partial W}{\partial S^a}$ equations require $\bar{F}^a F = 0$. By $SU(n)$ transformations, we can take

$$F = (a, 0, \ldots, 0), \quad D_{\bar{F}} = -\text{diag}(0, |b_1|^2, \ldots, |b_{n-3}|^2, 0, 0), \tag{2.13}$$

where $D_{\bar{F}}$ denotes the contribution to the $SU(n)$ D term from the $\bar{F}$'s. But, since the eigenvalues of the contribution to the D term from $A$ are all positive, there is no way to obtain a vanishing D term with $a \neq 0$. So we must require that the $\langle F \rangle = 0$. (ii) Now suppose that $A$ is non-zero. By an $SU(n)$ transformation,

$$A = \begin{pmatrix} a_1 \sigma_2 & & & \\ & a_2 \sigma_2 & & \\ & & \ldots & \\ & & & a_{n/2} \sigma_2 \end{pmatrix}. \tag{2.14}$$

This requires

$$D_{\bar{F}} = -\text{diag}(|b_1|^2, |b_1|^2, |b_2|^2, |b_2|^2, \ldots). \tag{2.15}$$

But the $\frac{\partial W}{\partial \bar{F}^a}$ equations, for general couplings, require that the $b_i$'s vanish. (To see this, take a special case: all $\gamma_{ij}$ vanish except $\gamma_{12}, \gamma_{23}, \ldots, \gamma_{n-5,n-4}$; this structure can be enforced by $U(1)$ symmetries, for example.) (iii) Finally, one can attempt to find a flat direction with $F$ and $\bar{F} = 0$. However, since $A$ and $S^a$ have the same *sign* of the $U(1)$ charge, this is impossible.

We will later analyze a specific model, and see that this class of theories opens up new possibilities for supersymmetry model building.

First, let us illustrate a few other possibilities. Consider a specific case: the $SU(7)$ model with an antisymmetric tensor and three $\bar{7}$'s. Now reduce the gauge group to $SU(5) \times SU(2) \times U(1)$, with the $U(1)$ generator taken to be $\tilde{T} = \text{diag}(2, 2, 2, 2, 2, -5, -5)$. The fields decompose as

$$A(10, 1, 4) + F(5, 2, -3) + S(1, 1, -10) + 3 \times \bar{F}^a(\bar{5}, 1, -2) + 3 \times \phi^a(1, 2, 5). \tag{2.16}$$

($a$ is a flavor index, and we suppress $SU(2)$ and $SU(5)$ indices.) Take for the superpotential:

$$W = \gamma A \bar{F}^1 \bar{F}^2 + \eta S \phi^1 \phi^2 + \lambda_a F \bar{F}^a \phi^a. \tag{2.17}$$

To see that there are no flat directions, we can proceed as in the $SU(n) \times U(1)$ example. First assume $F \neq 0$. Reasoning as above, this can be shown to be inconsistent. One then shows that $A$ and $\bar{F}^a$ must be zero. Finally, one must check that $\phi^a$ and $S$ vanish. If some $\phi^a$ is non-zero, then $S$ must be non-zero in order to insure vanishing of the $U(1)$ D term. The F-flatness condition, $\frac{\partial W}{\partial \phi^{1,2}} = 0$, requires both $\phi^1$ and $\phi^2$ to be zero. However, if both $\phi^1$ and $\phi^2$ vanish, the $SU(2)$ D term cannot vanish. Again, this model has a non-anomalous R symmetry and supersymmetry is broken.

Clearly this construction can be generalized in many ways, e.g. by reducing the $SU(2n+1)$ theory with antisymmetric tensor and $2n-3$ antifundamentals to $SU(2n-1) \times SU(2) \times U(1)$. Further examples are given in the Appendix.

## 2.3. A Non-Renormalizable Class of Models

Here we discuss a class of calculable models where supersymmetry breaking occurs along a D flat direction which is stabilized by a nonrenormalizable term in the superpotential. A simple model in which flat directions are lifted by non-renormalizable terms has gauge group $SU(6) \times U(1) \times U(1)_m$, where $U(1)_m$ is irrelevant for supersymmetry breaking but could play the role of messenger hypercharge, with D term automatically vanishing. The model possesses chiral superfields with the quantum numbers:

$$A(15,1,0) \quad \bar{F}^{\pm}(\bar{6},-2,\pm 1) \quad S^0(1,3,0) \quad S^{\pm}(1,3,\pm 2). \tag{2.18}$$

The gauge symmetries forbid a cubic superpotential in the model. At the level of dimension five terms, the unique allowed superpotential is:

$$W = \frac{\lambda}{M} A \bar{F}^+ \bar{F}^- S^0. \tag{2.19}$$

At this level, if one ignores the $U(1)_m$ symmetry, the model has a global $SU(2)$ symmetry (under which the $U(1)_m$ vector multiplet transforms like the $T_3$ generator).

To analyze the model, consider first the theory in the absence of the superpotential. There are then flat directions of the form

$$A = \begin{pmatrix} a\sigma_2 & 0 & 0 \\ 0 & b\sigma_2 & 0 \\ 0 & 0 & b\sigma_2 \end{pmatrix} \quad \bar{F} = \begin{pmatrix} c & 0 \\ 0 & c \\ 0 & 0 \\ 0 & 0 \\ 0 & 0 \\ 0 & 0 \end{pmatrix} \quad S^0 = d \quad S^{\pm} = e^{\pm}. \tag{2.20}$$

Here

$$2|a|^2 - |c|^2 = 2|b|^2, \qquad (2.21)$$

$$2|a|^2 - 4|c|^2 + 4|b|^2 + 3|d|^2 + 3|e^+|^2 + 3|e^-|^2 = 0. \qquad (2.22)$$

The first of these conditions is required by vanishing of the $SU(6)$ D term, the second by the $U(1)$ D term. The $U(1)_m$ D term vanishes for $e^+ = e^-$.

The gauge symmetry is broken to $Sp(4)$ in this direction. Gluino condensation in $Sp(4)$ leads to a non-perturbative superpotential. The form of this superpotential follows uniquely from symmetry considerations alone:

$$W_{np} = \frac{\Lambda^5}{\mathcal{O}^{1/3}}, \qquad (2.23)$$
$$\mathcal{O} = \bar{F}_i^+ \bar{F}_j^- A^{ij} \epsilon_{klmnop} A^{kl} A^{mn} A^{op}.$$

Turning on the non-renormalizable superpotential lifts the flat directions. We can ask how the vev's of the fields scale with the large scale, $M$. We will assume that $a \sim b \sim c \sim d$, and study of the potential shows that $e^\pm = 0$, so that messenger hypercharge is unbroken. Rescaling $(a, b, c, d) = M^{1/6} \Lambda^{5/6}(a', b', c', d')$, the potential has the form

$$V = \frac{\Lambda^5}{M} f(a', b', c', d'). \qquad (2.24)$$

In other words, the order of the vev's, the energy at the minimum $V_0$ and the Goldstino decay constant $F_G$ are

$$a, b, c, d \sim \Lambda^{5/6} M^{1/6}, \quad V_0 \sim \frac{\Lambda^5}{M}, \quad F_G \sim \frac{\Lambda^{5/2}}{M^{1/2}}. \qquad (2.25)$$

At the minimum of the potential, we expect that the expectation values of the auxiliary D fields are of order $F_G$. As a result, there is a non-supersymmetric contribution to the masses of the light fields $S^\pm$. Loop contributions involving messenger hypercharge bosons and the $S^\pm$ fields will lead to soft susy breaking masses for ordinary fields (along lines discussed in the next section). $\Lambda$, in this case, will be $10^2 - 10^3$ times larger than in the renormalizable case.

This model can be generalized to $SU(N) \times U(1)$ with $N > 4$ as follows. Take chiral matter superfields which transform under the gauge group plus a global $SU(N-4)$ as:

$$(A, N-4, 1) \qquad (\bar{F}, -(N-2), \overline{N-4}) \qquad (1, N, S_{ab}) \qquad (2.26)$$

where $S$ is a symmetric tensor of the global group. To stabilize the flat directions, the global $SU(N-4)$ must be explicitly broken down to a subgroup by the superpotential, but it is convenient to label fields by their $SU(N-4)$ content. One can gauge an anomaly free subgroup of this group (*e.g.* $SU(\frac{N}{2}-2)$ for $N$ even) to play the role of the messenger group. With a suitable superpotential, $W \sim A\bar{F}\bar{F}S$, there are no flat directions. Again, there is a non-anomalous R symmetry, and supersymmetry is broken.

In this paper we will not explicitly construct any realistic models with a nonrenormalizable supersymmetry breaking sector. However, this should be a straightforward exercise. One feature of such a model would be that the scale of R symmetry breaking is much higher than the scale of supersymmetry breaking, and so the properties of the R-axion are quite different. It might even be possible in some model to arrange for the R symmetry to be explicitly broken only by the QCD anomaly and for the R axion to be a phenomenologically acceptable QCD axion which solves the strong CP problem. (The superpotential would still have to be fine-tuned to make the cosmological constant zero, but perhaps this tuning does not require explicit R symmetry violation.)

### 2.4. A Model With Vanishing D Term

One of the main difficulties in the work of ref. [16] was the appearance of a Fayet-Iliopoulos D term for messenger hypercharge. This D term led to an undesirable pattern of symmetry breaking unless certain couplings were taken to be very small. Among the models we have developed here are some with discrete symmetries which, if unbroken, forbid a D term. This permits the construction of a much more compelling set of models.

An example of this phenomenon is provided by the renormalizable $SU(6) \times U(1) \times U(1)_m$ model. (The $U(1)_m$ symmetry plays the role of messenger hypercharge.) This model consists of the following representations:

$$A_{+2,0} \quad F_{-5,0} \quad \bar{F}^{\pm}_{-1,\pm 1} \quad \bar{F}^{0}_{-1,0} \quad S^{\pm}_{+6,\pm 1} \quad S^{0}_{+6,0}. \tag{2.27}$$

($A = 15$, $F = 6$, $\bar{F} = \bar{6}$ and $S = 1$ of $SU(6)$.) For the superpotential we take

$$W = \lambda A \bar{F}^{+} \bar{F}^{-} + \gamma F(\bar{F}^{+} S^{-} + \bar{F}^{-} S^{+}) + \eta F \bar{F}^{0} S^{0}. \tag{2.28}$$

Here we have imposed a discrete symmetry,

$$A \to A, \quad F \to +iF,$$
$$\bar{F}^{\pm} \to -i\bar{F}^{\mp}, \quad \bar{F}^0 \to -i\bar{F}^0, \quad (2.29)$$
$$S^{\pm} \to S^{\mp}, \quad S^0 \to S^0,$$

under which the $U(1)_m$ gauge fields change sign.

The following is a flat direction for $\lambda = 0$:

$$A = \begin{pmatrix} \frac{v}{\sqrt{2}}\sigma_2 & & \\ & 0 & \\ & & 0 \end{pmatrix}, \quad \bar{F}^- = \begin{pmatrix} v \\ 0 \\ 0 \\ 0 \\ 0 \\ 0 \end{pmatrix}, \quad \bar{F}^+ = \begin{pmatrix} 0 \\ v \\ 0 \\ 0 \\ 0 \\ 0 \end{pmatrix}. \quad (2.30)$$

In this flat direction, the original $SU(6) \times U(1) \times U(1)$ is broken to $SU(4) \times U(1) \times U(1)$. The low energy content of this model is exactly that of the $SU(4) \times U(1)$ model, plus three additional fields neutral under $SU(4)$ and the first $U(1)$. Among these are two fields carrying messenger hypercharge. These originate from the first two components of $\bar{F}^0$, and we denote them by $\chi^+$ and $\chi^-$.

Recalling the dynamics of the $SU(4) \times U(1)$ theory, we expect the vev's of the fields to have the following form:

$$A = \begin{pmatrix} \frac{v}{\sqrt{2}}\sigma_2 & & \\ & a\sigma_2 & \\ & & a\sigma_2 \end{pmatrix}, \quad S^0 = c,$$

$$\bar{F}^- = \begin{pmatrix} v \\ 0 \\ 0 \\ 0 \\ 0 \\ 0 \end{pmatrix}, \quad \bar{F}^+ = \begin{pmatrix} 0 \\ v \\ 0 \\ 0 \\ 0 \\ 0 \end{pmatrix}, \quad \bar{F}^0 = \begin{pmatrix} \chi^+ \\ \chi^- \\ b \\ 0 \\ 0 \\ 0 \end{pmatrix}, \quad F^0 = \begin{pmatrix} 0 \\ 0 \\ b \\ 0 \\ 0 \\ 0 \end{pmatrix}, \quad (2.31)$$

and all other vev's vanish.

We first ask whether messenger hypercharge is broken, *i.e.* whether the fields $\chi^{\pm}$ have non-vanishing expectation values. To analyze this problem, we consider the effective action at scales well below $v$. Integrating out the massive fields does not lead to superpotential couplings of the $\chi$ fields to the fields in the $SU(4) \times U(1)$ sector. On the other hand, integrating out massive gauge bosons at tree level leads to terms in the effective action of the form

$$\mathcal{L}_\chi = -\frac{1}{v^2} \int d^4\theta \ (\chi^{+\dagger}\chi^+ + \chi^{-\dagger}\chi^-)Z^\dagger Z \tag{2.32}$$

where $Z$ denotes some field with a non-zero F-component, such as the 4, $\bar{4}$ and antisymmetric tensor, $A$, of the low energy $SU(4)$ theory. Replacing these field by their expectation values yields mass terms for the scalar components of $\chi^\pm$. There are actually two types of gauge fields which contribute to these terms, in the limit that the $U(1)$ couplings are small compared to $g$, the $SU(6)$ coupling. These are associated with the broken generators,

$$\tilde{T} = \frac{1}{\sqrt{24}} \mathrm{diag}(2, 2, -1, -1, -1, -1) \tag{2.33}$$

and two sets of generators transforming in the 4 and $\bar{4}$ of $SU(4)$[1]. The masses of these fields are, respectively, $\frac{1}{2}g^2v^2$ and $\frac{3}{4}g^2v^2$. After a simple computation one obtains:

$$\mathcal{L}_\chi = -\frac{1}{v^2} \int d^4\theta \ (\chi^{+\dagger}\chi^+ + \chi^{-\dagger}\chi^-) \left(\frac{1}{6}A^\dagger A + \frac{2}{3}\bar{4}^\dagger \bar{4}\right). \tag{2.34}$$

This gives rise to a *positive* mass-squared for the scalar $\chi^\pm$ fields, so the symmetry is unbroken. Masses for the fermionic components of these multiplets are generated at one loop.

We wish to show that this vacuum leaves over a discrete symmetry under which the "messenger hypercharge" gauge boson is odd. Consider the transformation (2.29). This is, of course, not an invariance of the vacuum. However a combination of (2.29) with the $SU(6)$ transformation,

$$U = \begin{pmatrix} -i\sigma_1 & 0 & 0 \\ 0 & -i\sigma_3 & 0 \\ 0 & 0 & -i\sigma_1 \end{pmatrix}, \tag{2.35}$$

is unbroken. So there can be no D term.

---

[1] The reader trying to reproduce this computation may find it helpful to note that the model, as it stands, possesses an approximate, unbroken $SU(2)$ global symmetry which can be used to classify the generators

## 3. Model Building

*3.1. The Role of Messenger Hypercharge*

Our basic strategy for building models is close to that of ref. [16]. We will take one of the supersymmetry-breaking models described in the previous section, and gauge a global symmetry. This gauge interaction will serve as the messenger of supersymmetry breaking. It is tempting to take as messenger $SU(3) \times SU(2) \times U(1)$, but in all known cases, this requires a very large supersymmetry breaking group and yields a theory in which $SU(3)$ is not asymptotically free (we will comment on the possibility of exploiting recent developments to circumvent this problem in the conclusions). Instead, we will simply gauge a $U(1)$. It would be simplest to identify this $U(1)$ with ordinary hypercharge, or with another symmetry such as $B - L$ carried by ordinary particles. Again, however, there is a fundamental difficulty. Squarks and sleptons could all get mass-squared at two loops in this model, and the "bino" could get a mass at one loop. However, mass for the gluino would arise only at three loop order, and thus would be *extremely* small.

Instead, the messenger can be a $U(1)$ carried by hidden sector fields and some other, new fields. These new fields fall in vector-like representations of the standard model group. The SUSY breaking dynamics gives rise to multi-TeV masses for these fields, and also substantial splittings within the supermultiplets. Radiative corrections then lead to masses for squarks, sleptons and gauginos of a comparable order of magnitude.

Let us describe a particular model in some detail. We take, for the hidden sector, the $SU(6) \times U(1)$ model of the previous section. We take for the messenger group the $U(1)_m$ described there which has vanishing D term. In addition to these fields and the fields of the MSSM, we include a singlet $X$, two fields $\phi^+$ and $\phi^-$ with charge $\pm 1$, and an additional vector-like quark and lepton fields, $q$, $\bar{q}$, $\ell$ and $\bar{\ell}$. For this set of fields we take the superpotential to be

$$W_X = k_1 \phi^+ \phi^- X + \frac{1}{3} \lambda X^3 + k_3 X \bar{\ell} \ell + k_4 X \bar{q} q \ . \tag{3.1}$$

At two loops, the scalar components of $\phi^+$ and $\phi^-$ gain mass. The required calculation is quite straightforward, and very similar to that of the squark and slepton masses of ref.

[15]. For a range of parameters, this mass squared is negative:

$$m_\phi^2 = -\frac{1}{2}\left(\frac{\alpha_m}{\pi}\right)^2 m_\chi^2 \ln(\Lambda_6^2/m_\chi^2). \qquad (3.2)$$

Here $\Lambda_6$ is the scale of the $SU(6)$ theory; it is roughly the scale where the $\chi$ mass is determined.

As a result, the effective potential for $\phi^\pm$ and $X$ has the form, ignoring for a moment the terms involving $q$ and $\bar{q}$,

$$m_\phi^2\left(\left|\phi^+\right|^2 + \left|\phi^-\right|^2\right) + \left|k_1 X \phi^+\right|^2 + \left|k_1 X \phi^-\right|^2 + \left|k_1 \phi^+ \phi^- + \lambda X^2\right|^2 . \qquad (3.3)$$

At the minimum of this potential, $\phi^+$, $\phi^-$, $X$ and $F_X$ have non-zero vev's. For sufficiently small $\lambda$, this point is a minimum with zero vev's for the fields $q$, $\bar{q}$, $\ell$ and $\bar{\ell}$. Note that had there been a Fayet-Iliopoulos term at one loop for $U(1)_m$, $F_X$ would not have obtained a vev. This vev is crucial to what follows.

We can now consider loop contributions to the masses of squarks, sleptons and gauginos. These arise when we integrate out the fields $q$, $\bar{q}$, $\ell$ and $\bar{\ell}$. At one loop, for small $\lambda$, we obtain (majorana) masses for the $SU(3)$, $SU(2)$ and $U(1)$ gauginos to lowest order in $F_X$:

$$m_{\lambda_i} = c_i \frac{\alpha_i}{4\pi} \Lambda , \qquad (3.4)$$

where $c_1 = \frac{5}{3}$, $c_2 = c_3 = 1$, and the parameter $\Lambda$,

$$\Lambda = \frac{F_X}{X}, \qquad (3.5)$$

sets the scale for *all* of the soft breakings in the low energy theory. Masses for the squarks and sleptons appear due to gauge interactions at two loops. They are given by

$$\tilde{m}^2 = 2\Lambda^2\left[C_3\left(\frac{\alpha_3}{4\pi}\right)^2 + C_2\left(\frac{\alpha_2}{4\pi}\right)^2 + \left(\frac{Y}{2}\right)^2\left(\frac{\alpha_1}{4\pi}\right)^2\right]. \qquad (3.6)$$

Here $C_3 = 4/3$ for color triplets and zero for singlets; $C_2 = 3/4$ for weak doublets and zero for singlets, and $Y$ is the ordinary hypercharge.

Note the structure of the theory at this level. Squarks are the most massive scalar fields, by roughly a factor of three compared to slepton and Higgs doublets. Slepton singlets are the lightest scalar fields, by still another factor of order three. Gluinos have masses comparable to squarks, while the Majorana component of the wino mass matrix is comparable to that of the doublets. Note also that the strict degeneracy of squarks and of sleptons of the same gauge quantum numbers is only broken by effects of order quark or lepton Yukawa couplings. We will see that experimental constraints give masses for squarks and gluinos in the $200 - 300$ GeV range. This means that $\Lambda \sim 10 \ TeV$. This is the scale of $X$ physics. The scale of the hidden sector $SU(6) \times U(1)$ physics is larger by a factor of order $\frac{(4\pi)^2 \sqrt{\lambda}}{\alpha_m k_1^2}$, about $10^3$ TeV for coupling constants of order one. Note that this corresponds to a rather large value of the gravitino mass (of order 1 keV), which is marginally consistent with the upper bounds on the energy of the universe [23]. If the gravitino mass comes out too large, a period of late inflation might solve this and other problems (see §6).

We are particularly interested in the potential for the Higgs field. In the next section, we will explain how a $\mu H_U H_D$ term in the superpotential can be naturally generated in this framework. Here we note, first, that a coupling in the superpotential:

$$W_{XH} = \lambda' X H_U H_D \tag{3.7}$$

leads to a soft-breaking term $m_{12}^2 H_U H_D$ in the *potential*. Here $\lambda'$ must be rather small, since these masses should be roughly of order $(\alpha_2/\pi)^2$. This smallness is natural, in the sense of 't Hooft, in that it can arise due to approximate discrete or continuous symmetries. Note that the corresponding contribution to the $\mu$ term, however, is *extremely* small, far too small to be of phenomenological significance. Finally, a negative mass for $H_U$ arises from loops with top squarks. This contribution, although of three loop order, is somewhat larger than the two loop contributions because it is proportional to the top squark mass squared. We obtain

$$m_{H_U}^2 - m_{H_D}^2 = -\frac{3}{4\pi^2} y_t^2 \tilde{m}_t^2 \ln\left(\frac{\alpha_3}{\pi}\right), \tag{3.8}$$

where $y_t = \frac{m_t}{v_2}$ and, from eqn. (3.6),

$$m_{H_D}^2 \approx \frac{3}{2}\left(\frac{\alpha_2}{4\pi}\right)^2 \Lambda^2. \tag{3.9}$$

The argument of the logarithm is the ratio of the high energy scale, roughly of order $\Lambda$, to the stop mass.

To summarize, at energies well below the scale $\Lambda$, the theory looks like the usual MSSM, but with well-defined predictions for the soft breaking terms. Indeed, all of the soft breakings among the light states are determined in terms of three parameters: $\Lambda$, $\mu$, and $m_{12}^2$ (we view the $t$ quark mass as known, and for definiteness take $m_t = 175$ GeV. In a future analysis we will allow for a range of $t$-quark mass values). Other supersymmetry breaking terms, such as trilinear scalar couplings, are also generated but are small. In the next section, we will discuss the superparticle spectrum in this parameter space. Here we note that for a broad range of parameters, all of the current phenomenological constraints are satisfied. If one imposes some modest fine tuning constraints, however, much of the remaining parameter space will be explored at LEP II.

*3.2. The $\mu$ Term*

At first sight, it seems unnatural in the present context to have a $\mu$ term in the low energy theory. After all, the scale of the supersymmetry breaking is determined dynamically, and it would seem odd that the scale of the $\mu$ term and that of the weak scale should coincide. Moreover, the various mechanisms which have been suggested for generating a $\mu$ term in the standard supergravity framework are not available here, since the supersymmetry-breaking F components are not terribly large. As a result, in ref. [16], models with a low energy structure more complicated than that of the MSSM were considered. It appears to be natural and possible to construct models along these lines, but they require not only additional singlet fields but also additional vector-like quark and lepton fields. Moreover, these models introduce several new arbitrary coupling constants which affect the weak scale spectrum.

However, Leurer *et al.*, in a different context, have suggested a $\mu$ term generation mechanism which can be relevant here as well [20]. Suppose that, in addition to the usual MSSM fields, there is another singlet, $S$. As a consequence of discrete symmetries, the coupling $SH_UH_D$ is forbidden. Instead, the $S$ superpotential has the form

$$\frac{1}{M^n}S^{n+1}H_UH_D + \frac{1}{M^m}S^{m+3}. \tag{3.10}$$

In models with supergravity as the messenger of supersymmetry breaking, there is also a soft breaking mass term for $S$ of order $m_{3/2}^2$. If this term is negative, then

$$\langle S \rangle \approx M \left( \frac{m_{3/2}}{M} \right)^{\frac{1}{m+1}} . \quad (3.11)$$

This gives a $\mu$ term,

$$\mu \approx m_{3/2}^{\frac{n+1}{m+1}} M^{\frac{m-n}{m+1}} . \quad (3.12)$$

So, for example, if $n = m$, $\mu$ is of order $m_{3/2}$. In other words, for any discrete symmetry under which $H_U H_D$ carries the same charge as $S^2$, the $\mu$ term is of the correct order of magnitude.

In the present context, this mechanism has to be modified somewhat. There are a variety of possible contributions to the potential for $S$. These include various higher dimension couplings which can drive $\langle S \rangle$. In particular, consider terms in the effective lagrangian of the form:

$$\frac{1}{M_p^2} \int d^4\theta X^\dagger X S^\dagger S + \int d^2\theta \left( \frac{1}{M_p^p} X S^{2+p} + \frac{1}{M_p^m} S^{m+3} + \frac{1}{M_p^n} S^{n+1} H_U H_D \right) . \quad (3.13)$$

The first and second terms can contribute effective negative curvature terms to the $S$ potential. For example, if $p = 2$, $m = 2$ and $n = 1$, then the $\mu$ term is of order $\sqrt{F_X}$ times powers of coupling constants.

Besides generating a $\mu$ term, this mechanism can also generate a "B term", *i.e.* the soft supersymmetry breaking term $m_{12}^2 H_u H_d$ in the Higgs potential. However, examination of the potential resulting from eq. (3.13) shows that, in this example, the resulting B term is much too small, and so we must rely on the mechanism of the previous section to generate the B term.

These various structures of the $S$ superpotential can be enforced by discrete symmetries. We have not explored the full space of all possible couplings. The main lesson we wish to draw is that it is indeed possible to arrange a $\mu$ term of the correct order in these models. The price is a light field in the low energy theory (*e.g.* in our example above, the mass is of order $10^{-5}$ GeV). This field is very weakly coupled to ordinary matter, but could play a significant role in cosmology. We will not fully explore the cosmological implications of such a field in this paper, but will save a few remarks for §6.

The natural candidate for a symmetry that gives the $S$ superpotential of the form (3.10) is a discrete abelian symmetry, $Z_{m+3}$, with $S$ carrying charge –1, and $H_U H_D$ carrying charge $n-m-2$. Note that this discrete symmetry could be a horizontal symmetry and play a role in explaining the smallness and hierarchy in the fermion parameters. The ratio of scales, $\epsilon \sim \langle S \rangle / M_p$, would be the small breaking parameter of the horizontal symmetry. Various fermion parameters depend on different powers of $\epsilon$ and the hierarchy is naturally induced. Alternatively, the horizontal symmetry could be of the form $Z_{m_1} \times Z_{m_2}$, where each of the factors $Z_{m_i}$ is broken by a different singlet $S_i$ and $\epsilon \sim \langle S_1 \rangle / \langle S_2 \rangle$. Realistic examples of both types (in the sense that the small parameter is of the order of the Cabibbo angle) were constructed in [20] and [24]. An area for future exploration is whether a similar discrete symmetry can predict $\mu$ and B terms of the correct order of magnitude and explain the structure of the quark and lepton mass matrices in the case of low energy supersymmetry breaking.

## 4. Soft Breaking Phenomenology

Let us consider the low energy spectrum of the model we have constructed in previous sections in more detail. As we have already mentioned, its particle content at low energies is exactly that of the MSSM. However there are additional restrictions. There are only three free parameters: $\mu$, $m_{12}^2$ and $\Lambda$. We can trade the latter two for $m_Z^2$ and $\tan\beta = \frac{v_2}{v_1}$. After $m_Z^2$ has been fixed to its physical value, all masses can be expressed in terms of two parameters. The tree level Higgs potential has the form:[2]

$$V = m_1^2 H_d^{i*} H_d^i + m_2^2 H_u^{i*} H_u^i - m_{12}^2 (\epsilon_{ij} H_d^i H_u^j + \text{h.c.}) \\ + \frac{1}{8}(g^2 + g'^2)(H_d^{i*} H_d^i - H_u^{i*} H_u^i)^2 + \frac{1}{2}g^2 |H_u^{i*} H_d^i|^2, \quad (4.1)$$

where, in the present case,

$$m_1^2 = m_{H_D}^2 + |\mu|^2, \quad m_2^2 = m_{H_U}^2 + |\mu|^2, \quad (4.2)$$

---

[2] We do not consider radiative corrections to the Higgs potential here because the most severe constraints come not from neutral Higgs masses but from the lightest slepton masses.

with $m^2_{H_D}$ and $m^2_{H_U}$ as in eqns. (3.8) and (3.9). At the minimum,

$$\begin{aligned} m^2_1 &= m^2_{12}\tan\beta - \frac{1}{2}m^2_Z\cos(2\beta), \\ m^2_2 &= m^2_{12}\cot\beta + \frac{1}{2}m^2_Z\cos(2\beta), \\ m^2_A &= m^2_{12}(\tan\beta + \cot\beta). \end{aligned} \quad (4.3)$$

It is conventional in MSSM to express all Higgs masses in terms of $m^2_A$ and $\tan\beta$. In our case we can rewrite all masses in terms of $\mu$ and $\tan\beta$. Using minimization conditions and the fact that all scalar masses (including negative contribution of the top squark to the up-type Higgs mass (eqn. (3.8))) depend on one parameter only, we find:

$$m^2_A = \frac{2c\mu^2 - m^2_Z\cos(2\beta)(1-c)}{\cos(2\beta)(1-c) + c}, \quad (4.4)$$

where

$$c \equiv \frac{m^2_{H_D} - m^2_{H_U}}{2m_{H^2_D}} = \frac{2}{3\pi^2}\left(\frac{\alpha_3}{\alpha_2}\right)^2 y^2_t \ln\left(\frac{\alpha_3}{\pi}\right). \quad (4.5)$$

Note that $y_t$ here is determined in terms of $m_t$ and $\tan\beta$.

We can now use experimental bounds on the masses of the lightest Higgs and sleptons to constrain values of the parameters. Of these, the $SU(2)$-singlet sleptons provide the most severe constraint. Another stringent limit arises from considering bounds on charged Higgs mass from the rate of $b \to s\gamma$ [25]. There is no appreciable cancellation of the charged Higgs contribution by chargino loops. Examining the results of ref. [25], one finds that the charged Higgs must be rather massive and, correspondingly, in these models $\mu \geq 150$ GeV. To restrict the $\mu$ range from above we impose a fine tuning condition along the lines of ref. [26]:

$$\left|\frac{\mu}{m^2_Z}\frac{\partial m^2_Z}{\partial \mu}\right| \leq \Delta. \quad (4.6)$$

If we allow fine tuning of no more than 1 part in 10 ($\Delta = 10$) then $\mu \leq 200$ GeV. For $\Delta = 100$, $\mu \leq 600\ GeV$. These constraints (without inclusion of radiative corrections) are summarized in figure 1.

If we now take into account constraints from the neutralino sector (the lightest neutralino should be heavier than 25 GeV), $\mu \geq 200$ GeV, if positive, or $\tan\beta$ must be large (10 or larger). No additional restrictions for negative values of $\mu$ arise.

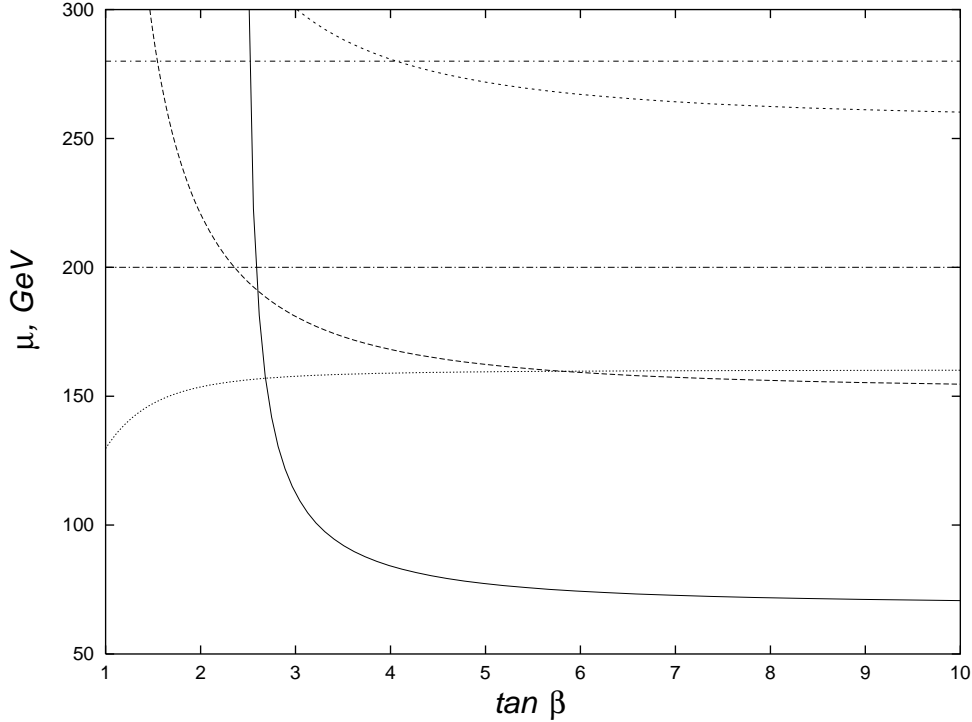

**Fig. 1.** Experimental constraints exlude regions below corresponding lines, fine tuning constraints exclude regions above corresponding lines. Solid line corresponds to light neutral Higgs boson mass (at tree level) of 65 GeV, dotted line corresponds to charged Higgs mass of 200 GeV, long dashed line corresponds to the selectron mass of 50 GeV, short dashed line shows region which will be covered by LEP 2 (selectron mass up to 80 GeV). Dash-dotted lines represent fine tuning constrains of 10%(lower line) and 5%(upper line).

For reasonable values of $\mu$ and $\tan\beta$, the masses of the $SU(2)$ singlet sleptons tend to lie between 50 and 65 GeV, so these particles should be discovered at LEPII. The lightest chargino has mass in the $50 - 85$ GeV range. Gluino masses tend to run from $225 - 300$ GeV, with squark masses somewhat larger. The lightest neutralino is in the range from $45 - 57$ GeV. So, unless one allows significant fine tuning, all of the masses tend to be on the small side.

These constraints will be relaxed in a non-minimal version of the model, with additional singlets, as in ref. [16]. Still, the minimal version is particularly simple and predictive.

## 5. Unification and the Sliding Singlet

One interesting feature of low energy supersymmetry breaking concerns the question of unification. The models we have described here are perturbatively unifiable (as far as $SU(3) \times SU(2) \times U(1)$ is concerned). In particular, all the fields we have added fall in complete $SU(5)$ multiplets. By itself, this is not particularly exciting. However, the most serious problem of conventional grand unified models is readily overcome in this framework: one can easily arrange that Higgs doublets are light while colored triplet fields are heavy. Most efforts to solve this problem use versions of the "missing partner mechanism" or similar group theoretic gymnastics. The resulting models typically involve enormous numbers of fields, and in some cases still suffer from potential fine tuning difficulties. An alternative approach, due to Witten [27], involves coupling a singlet field, $S$, to the Higgs field. If one simply examines the superpotential couplings, and studies the equation

$$\frac{\partial W}{\partial H_U} = (S + \mu)H_D = 0 \tag{5.1}$$

one seems to learn that either the doublet or the triplet fields are massless (here $\mu$ is a matrix with different entries for the doublets and triplets, typically due to the couplings to an adjoint field). In conventional susy breaking schemes, however, this mechanism is completely destroyed by terms in the Kahler potential which give rise to large tadpoles for $S$, of order $m^2_{3/2}M_p$ [28]. In contrast, in the present case, mass terms and tadpoles for $S$ are all of the order of the superpotential terms, and the mechanism can work [29]. The superpotential of the singlet can be flat enough if, for example, there is a discrete R symmetry under which $S$ is neutral. The presence of the singlet introduces some of the complications discussed in refs. [16,30], and probably requires additional fields. Still, this is possibly the most economical proposal within conventional grand unification for understanding this problem.

## 6. Some Cosmological Concerns

There are many cosmological issues raised by models of this type. We will not try and decide here whether a plausible model with acceptable cosmology exists. We would argue that the situation is similar to that of other supersymmetry and superstring cosmologies, where there are potentially serious problems and where solutions of varying degrees of plausibility have been suggested. Here, we will enumerate some of these issues.

1. Light gravitinos. In these models, the gravitino is light. Depending on how many couplings are required to communicate supersymmetry breaking to the ordinary sector and how large the coupling constants are, the gravitino mass ranges from less than an eV to over 10 keV. The longitudinal component is the goldstino, with an interaction strength about a million times smaller than ordinary weak interactions. If goldstinos are present at nucleosynthesis with a thermal density, they act as an additional neutrino species. This seems unacceptable (nucleosynthesis is currently in some trouble even with three light neutrinos). However, our gravitinos decouple in the early universe somewhat earlier than neutrinos do, before many particle species have decayed, and their abundance is diluted relative to the neutrino abundance by a factor of up to $\sim 100$, and so a mass as large as 10 keV is acceptable [23]. For higher masses, a period of late inflation could sufficiently dilute the gravitinos.

2. Domain walls. As has been discussed in refs. [15,16], there are typically discrete symmetries in these models, which can give rise to domain walls. One solution to this problem, suggested there, is that the discrete symmetries might be broken by dimension five operators, leading to collapse of the domain walls. Another solution is to find models where all the discrete symmetries have nonabelian gauge anomalies [31]. If the discrete symmetries are subgroups of spontaneously broken continuous symmetries, a remnant network of cosmological strings may remove the domain walls [32]. Still another alternative is that the domain walls might be diluted during a late period of inflation. Remember that the scales associated with the hidden sector are of order $10^5 - 10^7$ GeV or larger, *i.e.* high compared to the weak scale. Finally, we may be able to find models with no discrete symmetries.

3. Stable particles. The model we have presented predicts certain stable states (*e.g.* $q$, $\bar{q}$, $\ell$ and $\bar{\ell}$) which are potential dark matter candidates, since the remnant mass abundance of states with multi-TeV masses is typically comparable to closure density [33]. However, the existence of dark matter carrying standard model gauge quantum numbers is problematic [34]. The problem is worse if asymmetries in these particles are produced in the early universe. The most natural solution to these possible problems is that the heavy particles decay through higher dimension operators.

4. The moduli problem. If the underlying theory is a string theory, there could be moduli with very small masses. Some aspects of this situation have already been discussed in refs. [9,11].

5. Some of the fields we have introduced themselves behave in a manner similar to moduli. For example, the field $S$ which gave rise to the $\mu$ term is very weakly coupled. However, the characteristic energy contained in this field is not necessarily so large on cosmic scales. Its ultimate fate could well be tied with other moduli.

To summarize, we don't want to claim that the cosmological picture is rosy, but we see no insoluble cosmological problems.

## 7. Conclusions, or Where Do We Go From Here

Low energy supersymmetry breaking has, in principle, several attractive features when compared with more conventional supergravity-based models.

1. The hierarchy is readily explained in this framework.
2. It is highly predictive. Rather than involving 100 new parameters, typical models contain only a handful. In the models presented here, all of the soft breakings relevant to the MSSM were described in terms of two parameters.
3. Dangerous flavor-changing processes are automatically suppressed.
4. There is new physics (beyond that expected in the MSSM) at energy scales which might some day be accessible.

Here we have described models which achieve all of these goals. They are still somewhat complicated, but it is probably fair to say that they are not more complicated than any viable hidden sector supergravity model. More important, their complication no longer appears fundamental. No significant fine tuning is required in their construction. They are completely compatible with all phenomenological constraints. It seems reasonable to hope that, with a little more model building ingenuity, a more streamlined version of these ideas will emerge. Models without the intermediate stage of symmetry breaking connected with messenger $U(1)$ would be quite attractive. The tools which we have presented here – new theories exhibiting dynamical supersymmetry breaking, techniques for generating a $\mu$ term and eliminating Fayet-Iliopoulos terms – should be helpful in this process. Moreover, as Seiberg has suggested, it is possible that theories with difficulties such as Landau poles at low energies might be dual to theories without such problems [35].[3] Already, however, we believe that the low energy structure we have studied here is generic, and is likely to be true of any more "streamlined" model.

As we have discussed, the cosmology of these theories poses numerous challenges. This is also true, however, of models based on intermediate scale breaking. At the moment, only rather vague ideas exist as to how low energy supersymmetry breaking might be compatible with string theory or some other more fundamental structure. However, current string-based ideas also have serious problems with dilaton stability and the cosmological constant. Finally, we stress one of the great virtues of low energy based models: because they are predictive, experiment can definitively establish whether they are true.

## Acknowledgements

We thank Tom Banks, Howard Haber, Scott Thomas, Lisa Randall, Nathan Seiberg and Adam Schwimmer for conversations. The work of M.D. was supported in part by the U.S. Department of Energy. The work of A.N. was supported in part by the DOE under contract #DE-FG06-91-ER40614, and by a fellowship from the Sloan Foundation. The work of Y.N. is supported in part by the United States – Israel Binational Science Foundation (BSF), by the Israel Commission for Basic Research and by the Minerva Foundation.

## Appendix A. More Models of Dynamical Supersymmetry Breaking

*A.1. $SU(N) \times SU(2)$ Generalizations of the $SU(3) \times SU(2)$ Model*

The $SU(3) \times SU(2)$ model of ref. [1] can be generalized in several ways. An obvious generalization is to $SU(N) \times SU(2)$, $N$ odd, with chiral matter superfields transforming as $(N,2) + 2(\bar{N},1) + (1,2)$ and the analogous superpotential. A slightly less obvious generalization still has $N$ odd but takes for the matter sector

$$A \sim (A,1), \quad N \sim (N,2), \quad \bar{N}_i \sim (\bar{N},1), \quad D \sim (1,2) , \tag{A.1}$$

where $i = 1, ..., N-2$ is a flavor index. The superpotential is

$$W = \sum_{i,j=1}^{N-3} \gamma_{ij} A \bar{N}_i \bar{N}_j + \lambda \bar{N}_{N-2} N D . \tag{A.2}$$

---

[3] We thank R. Leigh and M. Strassler for a discussion of this point.

It is simplest to analyze this model by using the following gauge invariant chiral polynomials to parametrize the D-flat directions (see ref. [1] for some information on this technique):

$$X_{ij} = A \bar{N}_i \bar{N}_j, \quad Y_i = \bar{N}_i N D, \quad \Delta_{ij} = N N \bar{N}_i \bar{N}_j . \tag{A.3}$$

For generic superpotential, the equation $\frac{\partial W}{\partial \bar{N}_{N-2}} = 0$ forces $Y_i$ to be zero, $\frac{\partial W}{\partial A} = 0$ implies $\Delta_{ij} = 0$ and $\frac{\partial W}{\partial \bar{N}_i} = 0$, $i = 1, \ldots, N-3$, sets $X_{ij}$ to zero. Thus there are no flat directions. There is an R symmetry and the $SU(N)$ dynamically generates a superpotential so supersymmetry is presumably broken.

*A.2. A Nonrenormalizable $SU(N) \times Sp(M)$ Generalization of the $SU(3) \times SU(2)$ Model*

A rather clumsy generalization of the $SU(3) \times SU(2)$ model has gauge group $SU(N) \times Sp(M)$, $N$ odd, $M$ even, and $N > M$. Again this model has an R symmetry and $SU(N)$ dynamically generates a superpotential, so supersymmetry is broken if there are no flat directions. The matter fields are

$$Q \sim (N,M), \quad \bar{Q}_i \sim (\bar{N},1), \quad M \sim (1,M) , \tag{A.4}$$

where $i = 1, ..., M$ is a flavor index. The superpotential is

$$W = \lambda \bar{Q}_{M-1} Q M + \sum_{i,j=1}^{M-2} \gamma_{ij} Q Q \bar{Q}_i \bar{Q}_j . \tag{A.5}$$

To demonstate the absence of flat directions, note that, if $M$ is non zero, the $Sp(M)$ D terms require that $Q$ is non zero, and the condition $\frac{\partial W}{\partial \bar{Q}_{M-1}} = 0$ must be violated. Hence $M=0$ classically, and the $Sp(M)$ D terms require $Q$ to be of even rank. Now if some $Q^a$ is nonzero, $SU(N)$ D term cancellation requires some $\bar{Q}_i$ to also be nonzero. Then the conditions $\frac{\partial W}{\partial \bar{Q}_i} = 0$ ($i = 1, ..., M-2$) imply that the nonzero $\bar{Q}_i$ can only be $\bar{Q}_{M-1}$ or $\bar{Q}_M$. The condition $\frac{\partial W}{\partial M} = 0$ (combined with the $SU(N)$ D term condition) forbids $\bar{Q}_{M-1}$ to be non zero. Since we need an even number of the $\bar{Q}_i$ to be nonzero, they must all be zero. Hence there are no classical flat directions. In these models it is possible to choose $W$ to preserve a global $Sp(M-2)$ of which a subgroup may be gauged as messenger group.

### A.3. Another Class of Nonrenormalizable $SU(N) \times U(1)$ Models

Another infinite set of nonrenormalizable theories can be constructed as follows. Take the gauge group to be $SU(N) \times U(1)$, with $N > 3$. We have the option of preserving a global $SU(N-3)$ symmetry, although this is not necessary, and for convenience we group our superfields into $SU(N-3)$ multiplets. Our chiral superfields and their transformation properties under $SU(N) \times U(1) \times SU(N-3)$ are:

$$\begin{aligned} A &\sim (A, 2-N, 1), \quad N \sim (N, 1, 1), \quad \bar{N}_i \sim (\bar{N}, N-1, \overline{N-3}), \\ S_i &\sim (1, -N, N-3), \quad S_{ij} \sim (1, -N, A) , \end{aligned} \quad (A.6)$$

where $A$ stands for antisymmetric tensor and $i, j = 1, \ldots, N-3$. We take for the superpotential

$$W = \lambda_i \bar{N}_i N S_i + \gamma_{ij} A \bar{N}_i \bar{N}_j S_{ij}. \qquad (A.7)$$

Note that for $N = 4$, $S_{ij}$ does not exist and this is just the $SU(4) \times U(1)$ model.

It is not difficult to see that there is no flat direction here. Diagonalize $A^\dagger A$. The eigenvalues of this matrix are paired. For $N$ even, to obtain zero $SU(N)$ D term with $N^a \neq 0$, one needs some $\bar{N}_i^a \neq 0$. This is forbidden by the $\frac{\partial W}{\partial S_i} = 0$ equation, so $N = 0$. The $SU(N)$ D term conditions then require the rank of the $\bar{N}_i^a$ matrix to be even and, when combined with the $\frac{\partial W}{\partial S_{ij}} = 0$ equations, require all the $\bar{N}$ terms to vanish. The F terms and $SU(N)$ D terms allow only $A$, $S_i$ and $S_{ij}$ to be non-vanishing. However, with $N$ and $\bar{N}_i$ zero, the $U(1)$ D term can only vanish if $A$, $S_i$ and $S_{ij}$ are also zero, so there is no flat direction. For $N$ odd, one can also show that $N^a \neq 0$ requires some $\bar{N}_i^a \neq 0$, violating the $S_i$ F term conditions. With $\bar{N}_i^a = 0$, one can choose $A$ and $N$ to make the $SU(N)$ D terms vanish, but then the $U(1)$ D terms cannot be made to vanish. Hence there are no flat directions for $N$ odd either.

These theories all possess a non-anomalous $U(1)_R$ symmetry and a nonperturbatively generated effective superpotential, as in the $SU(4) \times U(1)$ model, and so supersymmetry is expected to be broken.

## References

[1] I. Affleck, M. Dine, and N. Seiberg, Nucl. Phys. B256 (1985) 557.

[2] M. Dine and N. Seiberg, Phys. Lett. 162B (1985) 299.

[3] N.V. Krasnikov, Phys. Lett. 193B (1987) 37.

[4] T. Banks and M. Dine, Phys. Rev. D50 (1994) 7454, hep-th/9406132.

[5] Y. Nir and N. Seiberg, Phys. Lett. 309B (1993) 337, hep-ph/9304307.

[6] L.E. Ibanez and D. Lust, Nucl. Phys. B382 (1992) 305.

[7] V.S. Kaplunovsky and J. Louis, Phys. Lett. 306B (1993) 269; R. Barbieri, J. Louis and M. Moretti, Phys. Lett. 312B (1993) 451.

[8] P. Pouliot and N. Seiberg, Phys. Lett. B318 (1993) 169, hep-ph/9308363; D. B. Kaplan and M. Schmaltz, Phys. Rev. D49 (1994) 3741, hep-ph/9311281; M. Dine, A. Kagan and R. Leigh, Phys. Rev. D48 (1993) 4269.

[9] T. Banks, D.B. Kaplan, and A.E. Nelson, Phys. Rev. D49 (1994) 779, hep-ph/9308292; B. De Carlos, J.A. Casas, F. Quevedo and E. Roulet, Phys. Lett. 318B (1993) 447, hep-ph/9308325.

[10] L. Randall and S. Thomas, MIT preprint MIT-CTP-2331, hep-ph/9407248.

[11] T. Banks, M. Berkooz and P.J. Steinhardt, Rutgers preprint RU-94-92, hep-th/9501053; T. Banks, M. Berkooz, S.H. Shenker, G. Moore and P.J. Steinhardt, Rutgers preprint RU-94-93, hep-th/9503114.

[12] M. Dine and D. MacIntire, Phys. Rev. D46 (1992) 2594, hep-ph/9205227.

[13] A.E. Nelson and N. Seiberg, Nucl. Phys. B416 (1994) 46, hep-ph/9309299.

[14] J. Bagger, E. Poppitz, and L. Randall, Nucl. Phys. B426 (1994) 3, hep-ph/9405345.

[15] M. Dine and A. E. Nelson, Phys. Rev. D48 (1993) 1277, hep-ph/9303230.

[16] M. Dine, A. E. Nelson and Y. Shirman, Phys. Rev. D51 (1995) 1362, hep-ph/9408384.

[17] I. Affleck, M. Dine and N. Seiberg, Phys. Rev. Lett. 52 (1984) 1677.

[18] D. Amati, K. Konishi, Y. Meurice, G.C. Rossi and G. Veneziano, Phys. Rep. 162 (1988) 169.

[19] K. Intriligator, N. Seiberg and S.H. Shenker, Phys. Lett. 342B (1995) 152, hep-ph/9410203.

[20] M. Leurer, Y. Nir and N. Seiberg, Nucl. Phys. B420 (1994) 468.

[21] N. Seiberg, Phys. Lett. 318B (1993) 469, hep-ph/9309335.

[22] K. Intriligator, R.G. Leigh and N. Seiberg, Phys. Rev .D50 (1994) 1092, hep-th/9403198.

[23] H. Pagels and J.R. Primack, Phys. Rev. Lett. 48 (1982) 223.

[24] T. Banks, Y. Grossman, E. Nardi and Y. Nir, hep-ph/9505248.

[25] A.J. Buras and S. Pokorski, Nucl. Phys. B424 (1994) 374, hep-ph/9311345.

[26] R. Barbieri and G.F. Giudice, Nucl. Phys. B306 (1988) 63.

[27] E. Witten, Phys. Lett. 105B (1982) 267.

[28] J. Polchinski and L. Susskind, Phys. Rev. D26 (1982) 2661; M. Dine, Proceedings of the Johns Hopkins Workshop on Particle Physics (1982); T. Banks and V. Kaplunovsky, Nucl. Phys. B206 (1982) 45; H.P. Nilles, M. Srednicki and D. Wyler, Phys. Lett. 124B (1982) 337; A.B. Lahanas, Phys. Lett. 124B (1982) 341.

[29] D. Nemeschansky, Nucl. Phys. B234 (1984) 379.

[30] J. Bagnasco, Ph.D. Thesis (unpublished).

[31] J. Preskill, S.P. Trivedi, F. Wilczek and M.B. Wise, Nucl. Phys. B363 (1991) 207.

[32] T.W.B. Kibble, G. Lazarides and Q. Shafi, Phys. Rev. D26 (1982) 435; A. Vilenkin, Phys. Rep. 121 (1985) 263 and references therein.

[33] S. Dimopoulos, N. Tetradis, R. Esmailzadeh and L. J. Hall, Nucl. Phys. B349 (1991) 714, and references therein, ERRATUM-ibid. B357 (1991) 308.

[34] S. Dimopoulos, D. Eichler, R. Esmailzadeh and G. D. Starkman, Phys. Rev. D41 (1990) 2388; A. De Rujula, S.L. Glashow and U. Sarid, Nucl. Phys. B333 (1990) 173; A. Gould, B.T. Draine, R.W. Romani and S. Nussinov, Phys. Lett. 238B (1990) 337; J.L. Basdevant, R. Mochkovitch, J. Rich, M. Spiro and A. Vidal-Madjar, Phys. Lett. 234B (1990) 395; E. Nardi and E. Roulet, Phys. Lett. 245B (1990) 105.

[35] N. Seiberg, Nucl. Phys. B435 (1995) 129, hep-th/9411149.